\documentstyle[11pt,aaspp4]{article}
\input epsf

\accepted{August 11, 1998}
\journalid{337}{}
\articleid{11}{14}

\def\spose#1{\hbox to 0pt{#1\hss}}
\def\lta{\mathrel{\spose{\lower 3pt\hbox{$\mathchar"218$}}
     \raise 2.0pt\hbox{$\mathchar"13C$}}}
\def\gta{\mathrel{\spose{\lower 3pt\hbox{$\mathchar"218$}}
     \raise 2.0pt\hbox{$\mathchar"13E$}}}
\def\n{\noindent}

\def\be{\begin{equation}}
\def\ee{\end{equation}}
\def\msun{M_{\odot}}
\def\rsun{R_{\odot}}
\def\mearth{M_{\oplus}}

\begin{document}

\title{Planet Consumption and Stellar Metallicity Enhancements}
\author{Eric Sandquist\altaffilmark{1}, Ronald E. Taam\altaffilmark{1}, 
D. N. C. Lin\altaffilmark{2}, Andreas Burkert\altaffilmark{3} }

\n \altaffilmark{1}{Department of Physics \& Astronomy, Northwestern 
University, Evanston, IL 60208
\vskip -0.12in
erics@apollo.astro.nwu.edu, taam@ossenu.astro.nwu.edu}

\n \altaffilmark{2}{UCO/Lick Observatory, Board of Studies in Astronomy 
and Astrophysics, University of California, Santa Cruz, CA 95064
\vskip -0.12 in
lin@ucolick.org}

\n \altaffilmark{3}{Max Planck Institut for Astronomy, Koenigstuhl 17, D-69117
Heidelberg, Germany
\vskip -0.12in
burkert@mpia-hd.mpg.de}

\begin{abstract}

The evolution of a giant planet within the stellar envelope of a
main-sequence star is investigated as a possible mechanism for
enhancing the stellar metallicities of the parent stars of extrasolar
planetary systems.  Three-dimensional hydrodynamical simulations of a
planet subject to impacting stellar matter indicate that the envelope
of a Jupiter-like giant planet can be completely stripped in the outer
stellar convection zone of a $1 \msun$ star.  In contrast,
Jupiter-like and less massive Saturn-like giant planets are able to
survive through the base of the convection zone of a $1.22 \msun$
star.  Although strongly dependent on details of planetary interior
models, partial or total dissolution of giant planets can result in
significant enhancements in the metallicity of host stars with masses
in the range $1.0 \msun \lta M \lta 1.3 \msun$.  The implications of
these results with regard to planetary orbital migration are briefly
discussed.
\end{abstract}


\section{INTRODUCTION}

With the detection of planets orbiting around nearby solar-type stars
(Mayor \& Queloz 1995; Butler \& Marcy 1996; Marcy \& Butler 1998)
there has been renewed interest in the origin and orbital evolution of
planetary systems.  Particularly noteworthy about a large fraction of
the discoveries was the small separations (less than 0.2 AU) at which
the Jupiter-sized planets orbited about their parent stars.  Based on
our current understanding of the formation of the giant planets, it is
likely that they formed at large distances from their parent stars (on
the order of a few AU) where temperatures in the protoplanetary disk
were sufficiently low that refractory material could condense.
Subsequently the giant planets must have undergone significant inward
orbital migration either due to tidal interaction with the remnant
disk and star (Lin, Bodenheimer, \& Richardson 1996; Trilling et al.
1998; Murray et al. 1998) or to gravitational interactions between the
planets themselves (Rasio \& Ford 1996) during the remaining lifetime
of the system.

Investigations of the chemical composition of the stellar companions
in these systems (Gonzalez 1997, 1998a) indicate that the metallicity
of the stellar photospheres is enhanced above the solar value, which
is itself at the high end of the metallicity distribution of nearby
G-dwarfs (Favata, Micela, \& Sciortino 1997).  This suggests that
either metal-rich environments are more conducive to the formation of
these types of planetary systems, or that orbital migration has led to
the accretion of planetary bodies into the stellar envelope (Lin 1997;
Gonzalez 1997, 1998ab; Jeffrey, Bailey, \& Chambers 1997; Laughlin \&
Adams 1997) since gas giants in the solar system appear to have
average metal contents of Z $\gta 0.1$ (Gudkova \& Zharkov 1990). In
this interpretation, the metallicity enhancement is
restricted to their outermost envelope layers and is not indicative of
the star as a whole.

In this Letter we explore the accretion hypothesis and report on the
three-dimensional hydrodynamical evolution of a planet moving through
the envelope of its parent star.  The goals of this investigation are
the determination of the amount of mass removed from the planet due to
mass stripping and shock heating, and the consequent metallicity
enhancement expected for the star.  Provided
that the planetary envelope is at least partially removed and/or the
planetary core is melted and vaporized within the outer stellar
convection zone, the metallicity of the stellar envelope can be
enhanced.

\section{FORMULATION}

In this study it is assumed that the planet is in a circular orbit
about its parent (main sequence) star and, due to computational
resource limitations, is already partially immersed in the outermost
regions of the stellar envelope. The orbital evolution immediately
prior to this phase is governed by the action of gravitational tidal
torques with the star (Lin 1997). Low-mass main sequence stars brake
their rotation on a timescale of approximately 10$^8$ yr (Skumanich
1972) once the protoplanetary disk has been dissipated, allowing the
star to exert drag on the planet.  This spiral-in phase and the
possible mass loss from the planet associated with tidal effects has
not been followed, and it is beyond the scope of the present
investigation. (This scenario would occur later than the pre-main
sequence scenario considered by Laughlin \& Adams 1997).  The
structure of the host star was calculated from a stellar evolutionary
code developed by Eggleton (1971, 1972) and updated by Pols et al.
(1995). The star in the first simulation had a mass of $1
\msun$ and an age of $4.65 \times 10^9$~years, with a convection zone
containing $0.02 \msun$ and extending over the outer 28\% of its
radius. A $1.22 \msun$ model at an age of $10^9$~years (having an
outer convection zone of $0.003 \msun$ and a fractional radial extent
of 18\%) was used in the remaining two simulations.  For the planet,
we use polytropic models with an index $n$ equal to 1.0 or 1.5,
which are good approximations to the density structure of Jupiter and
Saturn, respectively (Hubbard 1984).

The three-dimensional hydrodynamical evolution of the planet within
the stellar envelope is calculated using a nested grid technique in a
code developed by Burkert \& Bodenheimer (1993) as described in
Sandquist et al. (1998). The computational domain is composed of two
grids of $64\times 64\times 64$ zones having total physical sizes of
$5.2 \times 10^{10}$~cm and $1.3 \times 10^{10}$~cm respectively.  The
subgrid was positioned over the center of the planet, and subsequently
we followed the planet's center of mass reference frame by imprinting
the orbital motion of the planet with respect to the stellar envelope
on the velocity field of the inflowing matter.  The subgrid was placed
off-center in the main grid in order to allow us to follow more of the
disturbed flow.

The orbit of the planet decays as the result of momentum transfer from
the stellar to the planetary envelope.  Gravitational focusing and
drag is unimportant in the present circumstance since the accretion
radius is more than a factor of 50 times smaller than the planet's
radius for the conditions under consideration here (viz., orbital
velocity of $4.4 \times 10^7$~cm s$^{-1}$ and a planetary mass and
radius comparable to that of Jupiter $\sim 0.001 \msun$ and $\sim 0.1
\rsun$).  The aerodynamic drag force acting on the planet's surface was
taken into account by computing the planet's center-of-mass velocity
relative to the grid at each step, and correcting the planet's orbital
motion accordingly. The instantaneous mass of the planet was computed
as the sum of zones in the subgrid for which the speed was less than
20\% of the orbital motion of the planet. The planet was recentered if
it moved more than 0.3 of a grid zone, thus ensuring adequate spatial
resolution.

A number of simplifying approximations have been introduced in the
problem to facilitate computation.  For example, we have implicitly
assumed that the planet is always encountering an inflowing density
distribution corresponding to the undisturbed stellar envelope.  This
approximation should be adequate during the planet's main mass loss
phase since the timescale for orbital evolution is shorter than the
orbital period during this phase.  On the other hand, this
approximation is suspect in the early evolution where the planet skims
the surface of the star, making orbits at approximately the same
radius.  However, very little mass is expected to be lost during this
stage due to the low densities in the stellar atmosphere, so the total
mass loss from the planet is less likely to be significantly affected.
Perhaps a more serious approximation is the lack of treatment of the
transfer of orbital angular momentum to spin angular momentum of the
stellar material.  Angular momentum exchange arises from tidal
interaction as well as mass transfer.  This results in an overestimate
of the relative velocity and an underestimate of the timescale for the
evolution of the planet within the stellar envelope. The effect on
mass loss from the planet should not be large since the region in
which the planet has significant gravitational influence on the
incoming stellar matter is much smaller than its own radius, and the
planet's interaction with disturbed stellar matter is not expected to
be significant. Finally, the gas in the stellar envelope and in the
planet is treated as a perfect gas, and the compositional differences
between the planetary material and the stellar envelope are neglected
in the simulations.  This latter approximation should not
significantly affect the overall results, however, since the energy
required to dissociate and ionize the gas within the planetary
envelope is much less than the energy input associated with the impact
of the stellar material.

\section{NUMERICAL RESULTS}

Three numerical simulations of a planet within the envelope of a main
sequence star have been performed. In the first sequence we examined
the impact of the stellar envelope in the $1 \msun$ star on a
Jupiter-like giant planet.  The initial structure of the planet was
described by an $n=1$ polytrope with mass and radius equal to the
mass, $M_J$, and radius, $R_J$ of Jupiter.  The planet was initially
placed in orbit at a radius of $0.98 \rsun$.

As illustrated in Fig. 1 the distance of the planet from the
center of the star rapidly decreased from $0.98 \rsun$ to $0.78 \rsun$ in
about 17000 s.  The mass of the planet only decreased by 11\% on the first
orbit (about 9900 s). As the planet plunged deeper into the stellar envelope
(for depths greater than $0.05 \rsun$ from the stellar surface --- see
Fig.  1), the mass loss from the planet increased dramatically.  In
this simulation, the planet lost 90\% of its mass once it has passed
through 80\% of the convection zone's radial extent.  The planet was
significantly distorted from its initial spherical distribution, most
rapidly on the side that was deeper in the stellar envelope. The
leading edge was also significantly compressed. The effect of shock
heating is clearly evident in Fig. 2 where the entropy distribution in
the flow is illustrated for two snapshots.  The regions of high
entropy in the flow identify regions that were shock heated. At the
end of the simulation, the stronger shock was contributing to
quicker mass loss from the planet.

\begin{figure}[p]
\hspace*{-0.3in}
\centerline{\epsffile{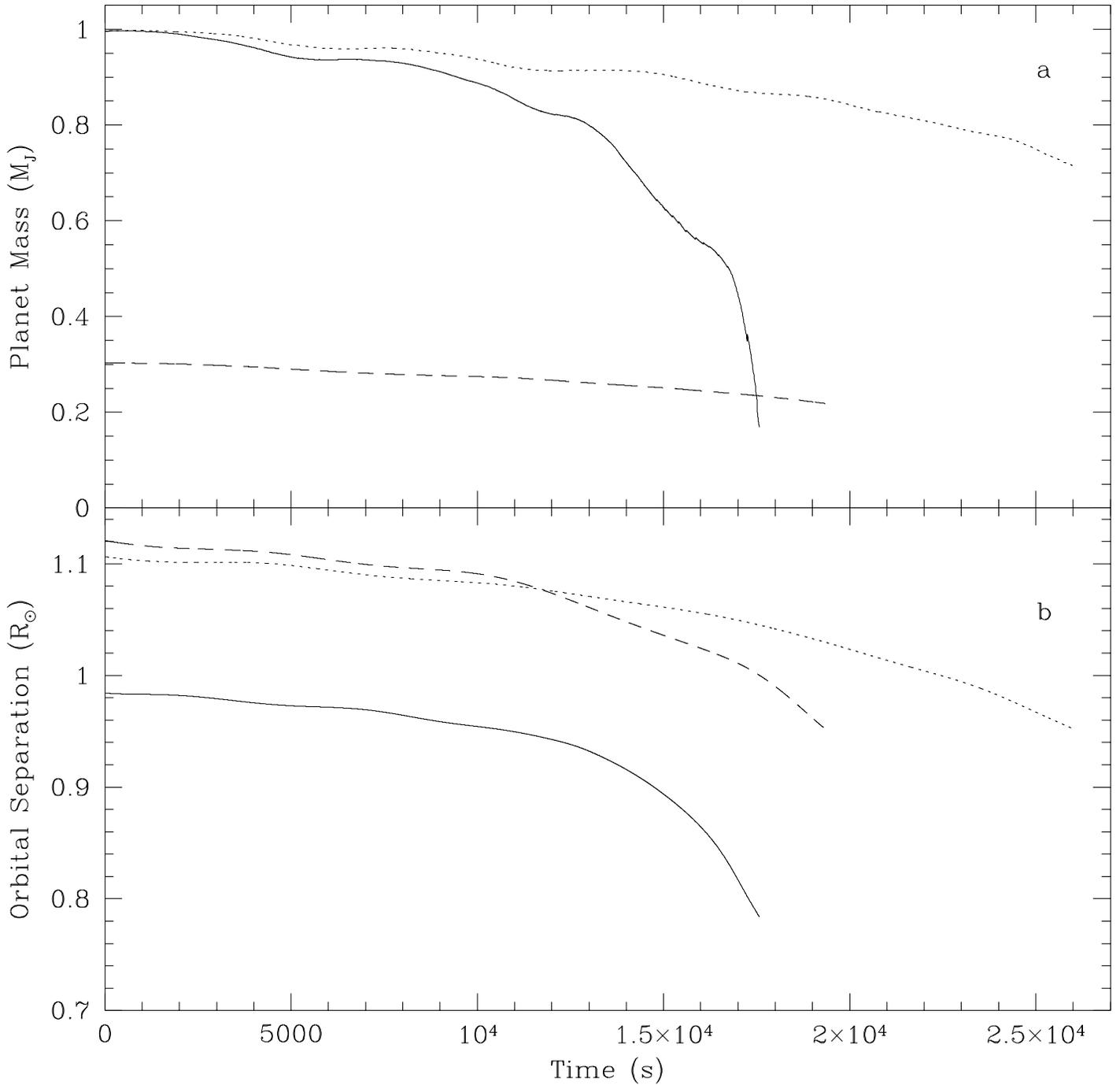}}
\figcaption{a) The temporal variation of the planet mass for
the Jupiter - $1 \msun$ star simulation ({\it solid line}), the
Jupiter - $1.22 \msun$ star simulation ({\it dotted line}), and the
Saturn - $1.22 \msun$ star simulation ({\it dashed line}).
b) The temporal variation of the  orbital separation for the three
simulations.}
\end{figure}
\begin{figure}
\centerline{\plotfiddle{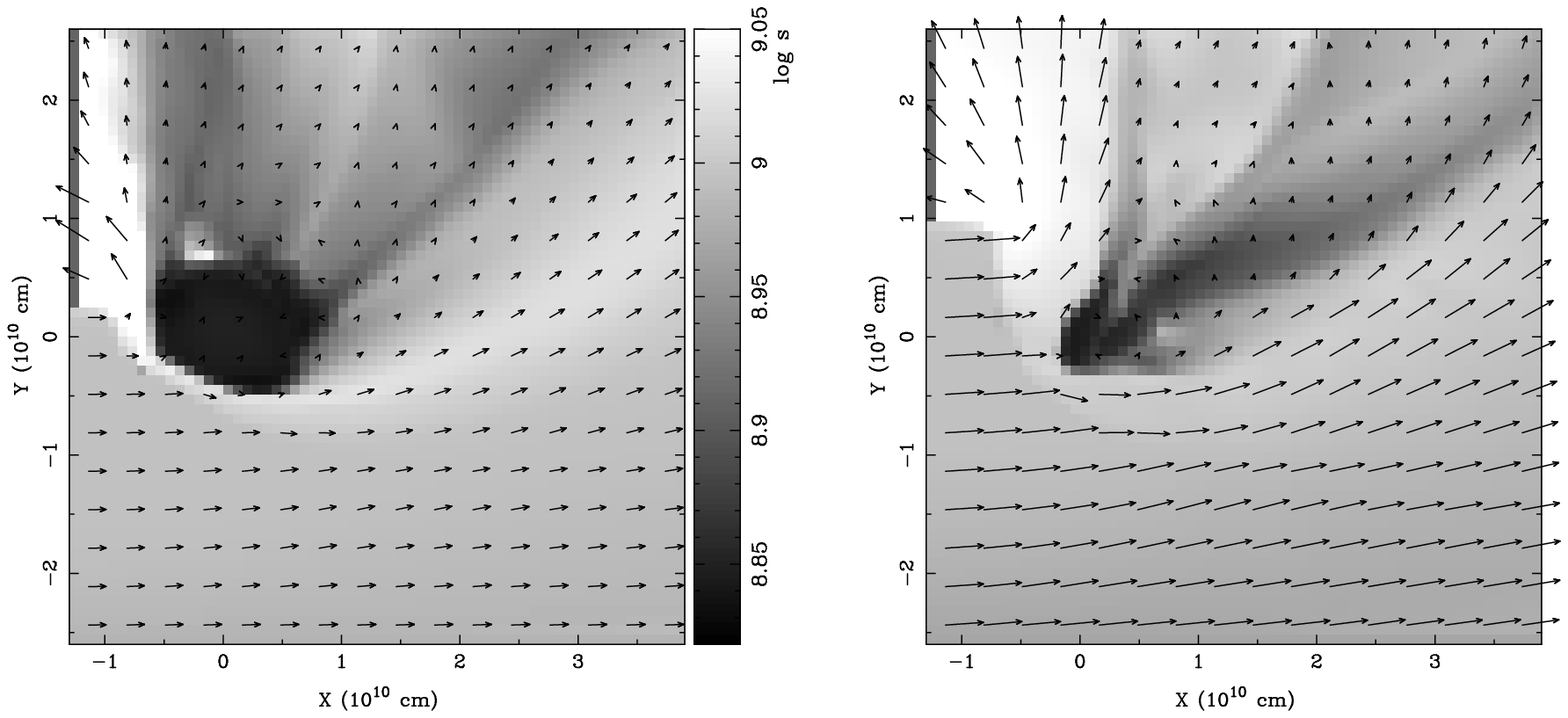}{20cm}{90}{100.0}{100.0}{432}{0}}
\figcaption{The entropy distribution and velocity field in the
meridional plane of the Jupiter-like planet in the envelope of a $1
\msun$ star at times a) 6740 s, and b) 15999 s.  The $x$- and $y$-axes
correspond to the orbital direction and the radial direction
respectively. The entropy of the original surface of the planet was
log $s = 8.91$.  The velocity vectors are scaled to the maximum
in-plane speed.  The maximum speeds for the panels are: a) 985 km
s$^{-1}$ (6740 s), and b) 468 km s$^{-1}$ (15999 s).}
\end{figure}

To determine the sensitivity of the results to the assumed mass of the
parent star, the $1 \msun$ star was replaced with a $1.22 \msun$ star
in the second and third simulations.  Because the orbital decay
timescale is longer in this case, we started the planet in orbit at
$0.96 R$. By the time the planet reached the base of the convection
zone, it had lost a smaller amount of mass ($\sim 0.3 M_J$) due to the
lower densities in the convection zone.  Although the orbital decay
timescale in this sequence is comparable to the orbital period of
the planet during the main mass loss phase (so that the planet is
likely to encounter stellar material disturbed by its previous
passage), the planet is still unlikely to dissolve in the convection
zone.  This is a consequence of the feedback between the amount of
momentum incident on the planet and the planet's rate of orbital
decay: a higher (lower) momentum flux results in a faster (slower)
decay.  As a result, a large change in the mass loss from the planet
is less likely.

In the final simulation, we sent a Saturn-like planet through the
envelope of a $1.22 \msun$ star. Here, the initial interior structure
of the planet is described by a polytropic index of 1.5 and the planet
is characterized by a mass and radius of 0.3 $M_J$ and 0.81 $R_J$
respectively.  Figure 1b illustrates that the planet's orbit decayed
more quickly than did Jupiter's. Because Saturn has a more
centrally-concentrated mass distribution than Jupiter, the steeper
density gradients were better able to inhibit the inward propagation
of shock waves, resulting in momentum transfer to the planet as a
whole, rather than mass loss. In agreement with this idea, our
Saturn-like planet lost less mass ($0.09 M_{J}$) than our Jupiter-like
planet did ($0.17 M_{J}$) after both had swept through the same amount
of mass in the stellar envelope ($0.23 M_{J}$).

\section{DISCUSSION}

Before discussing the potential metallicity enhancements that could be
observed as a result of star-planet interactions like the ones we have
simulated here, we must first summarize current beliefs regarding the
chemical composition and interior structure of the gas giants in the
solar system. Recent models (Guillot, Gautier, \& Hubbard 1997)
indicate that the total amount of elements heavier than helium in
Jupiter is between 11 and $45 \mearth$, with the preferred equation of
state implying a mass of less than $33 \mearth$. The interior distribution
of metals is not well-constrained --- there are no
observable differences if the metals are distributed uniformly though
the metallic hydrogen region, or are condensed into a dense
core. Though the core-instability model for giant planet formation
requires a core mass of about $15 \mearth$ to initiate rapid gas
accretion for Jupiter (Podolak et al. 1993, Pollack et al. 1996), the
mass of the core can decrease during the formation phase due to
heating by accreting material. Rock/ice cores of masses between 2 and
$12 \mearth$ appear to be necessary with Guillot et al.'s
preferred equation of state. Models of extrasolar planets (Burrows et
al. 1997) have hinted at the presence of radiative zones in planets
like Jupiter and Saturn.  This can be relevant if significant
heavy-element enrichment of the planet's envelope resulted from the
impact of rock/ice planetesimals after it had accreted the majority of
its gas.

From three-dimensional hydrodynamical simulations of the evolution of
a Jupiter-like planet in the convective envelope of a solar-mass star,
we found that the planetary envelope can be completely stripped
away. If the planet has a solid core, it is also likely to be melted
and vaporized in the convection zone since the rate of heating
associated with the impacting matter ($\sim \pi R^2 \rho v^3 \sim
10^{40}$ ergs s$^{-1}$) is sufficient to dissolve the core on a short
timescale ($\sim 10-100$ s). Thus, the dissolution of a single
Jupiter-like planet could lead to a heavy element enrichment of
between 3 and 26\% in the envelope of a star of solar mass and
metallicity.  Greater metallicity enhancements in solar-type stars
could be achieved if a few Jupiter-like planets or large numbers of
planetesimals were accreted. The likelihood of these kinds of
accretion events depends on a number of unknown factors like the mass
function for planets, the lifetime of the protoplanetary disk, and the
orbital history of the planets and residual debris. In determining the
metallicity enhancements, we have also assumed that there is no
elemental diffusion or mixing across the radiative-convective
boundary.  Efficient mixing across this boundary would result in some
depletion of light elements like lithium and beryllium (although the
planet itself would introduce a fresh supply), and would modify the
internal angular momentum distribution of the star.

The numerical simulation for a $1.22 \msun$ star reveals that a
Jupiter-like planet loses about 30\% of its envelope mass in the
stellar convection zone. The heavy-element enrichment of the star
could reach above 50\% if Jupiter's heavy elements are distributed
uniformly through its envelope. However, this is still subject to the
considerable uncertainty in the planet composition models --- no
enrichment would be observed if the heavy elements were concentrated
in Jupiter's core.  In the simulation using a Saturn-like planet, the
metallicity enhancement would be about 14\% at most, and the planet
model uncertainties are greater.  Models of Saturn imply a total metal
mass of up to $20 \mearth$ including a rock-ice core of approximately
$7 \mearth$ (Gudkova \& Zharkov 1990).

The envelope structure of the host star is clearly an important factor
in the metallicity enrichment by planets since less massive convection
zones could magnify the enhancements. However, main-sequence stars
more massive than the Sun have thinner and less dense convection
zones, and so they are not as effective in stripping the planetary
envelope. Meridional circulation, induced by the angular velocity
gradients resulting from the deposition of orbital angular momentum in
the stellar envelope, is not likely to be suppressed in the radiative
envelope of stars more massive than about $1.4 \msun$. Hence, the
metals are expected to be well mixed in a region of the stellar
envelope which is affected by rotational instabilities.  In a separate
paper, we will examine the extent of metal enhancement in these more
massive stars.  The deeper and higher density convection zones in
stars less massive than the Sun would be more effective in stripping,
but their greater mass more effectively dilutes the added metals.
Since the convection zone mass rapidly increases with decreasing
stellar mass, stars significantly less massive than the Sun are
unlikely to show metallicity enhancements even if they consume several
planets. Based on the calculations presented here, if a solar-mass
star consumed one Jupiter-like planet, its metallicity would only be
enhanced by at most 0.1 dex (if the star initially had solar
metallicity). Thus, we expect that metallicity enhancements can only
be observed in stars ranging from about $1.0 - 1.3 \msun$.

Gonzalez (1998a) has measured abundances and estimated masses for most
of the stars believed to have planetary systems. He finds that the
four ``51 Peg-like'' systems (51 Peg, $\upsilon$ And, $\tau$ Boo,
$\rho^1$ Cnc) have a mean abundance [Fe/H]$=+0.25$, and two ($\tau$
Boo and $\rho^1$ Cnc) appear to belong to the small group (about 8
members) of known ``super metal-rich'' stars (Taylor 1996).  The stars
with planet candidates were selected as being solar-type stars, so
the star sample does not tell us in a general sense whether stars with planets
fall in the range where we would predict that metallicity enhancements
are possible. However, of the four, $\upsilon$ And and $\tau$ Boo have mass
estimates of approximately $1.3 \msun$, 51 Peg has about $1.1 \msun$,
and $\rho^{1}$ Cnc may be as low at $0.7 \msun$ (Gonzalez 1998a, Ng \&
Bertelli 1998).

The orbital angular momentum deposited in the stellar convection zone
may influence the orbital migration of the remaining planets in the
system.  For example, if the surface layers of the central star are
spun up as a result of the angular momentum deposited in the envelope,
then the remaining planets within several stellar radii may spiral
outwards if the stellar spin frequency is greater than the orbital
frequency of the planet.  The angular momentum deposited in the
stellar surface layers may also facilitate dynamo action, enhancing
the magnetic field of the central star.  This can lead to enhanced
magnetic braking of the stellar surface, thereby spinning down the
star. Spectroscopic observations of stars with extrasolar planet
candidates show mixed evidence for higher spin rates, as $\tau$ Boo
and $\upsilon$ And have $v \sin i$ of about 10 km s$^{-1}$ (Gonzalez
1997), and all others have $v \sin i$ of just a few km s$^{-1}$
(Gonzalez 1998a).

Residual planetesimals, comets, and short-period terrestrial planets may also
contaminate the host stars with heavy elements.
In one scenario for the origin of short-period planets,
Murray et al. (1998) suggest that a giant planet can induce eccentricity
growth among residual planetesimals through resonant interactions.
Subsequent close encounters cause most of the affected planetesimals to be
ejected outwards while the planet migrates inward.  A
Jupiter-mass planet could only migrate a large distance inward if
there was a substantial population of planetesimals within its orbit.
At the distance of a few stellar radii, some fraction of the close
encounters would send the planetesimals toward their host stars to be
dissolved in the outermost layers of the stellar envelope.  The total
mass of planetesimals needed to explain all of the heavy elements
contained in the convective envelope of solar-type stars (a few tens
of Earth masses) is considerably smaller than what is required to
induce a Jupiter-mass to undergo any significant orbital decay.
Unless most of the acquired heavy elements are able to diffuse into
the radiative interior, the planetesimal-scattering scenario for
orbital migration would require more than 90\% of the close encounters
to result in the outward ejection of planetesimals.

\acknowledgements

\n We wish to acknowledge discussions with Prof. Peter Bodenheimer.
This research was supported by NASA under grant NAG5-7515 and by 
NSF grants AST-9415423, AST-9618548, and AST-9727875.

\end{document}